\documentclass{aa}
\usepackage{amsmath,amssymb,graphicx,natbib}
\bibpunct{(}{)}{;}{a}{}{,}
\def\bx{{\bf x}}
\def\by{{\bf y}}
\def\cN{{\mathcal N}}
\def\cO{{\mathcal O}}
\def\dt{{\!{\rm d}t\,}}
\def\du{{\!{\rm d}u\,}}
\def\ddx{{\!{\rm d}^dx\,}}
\def\ddy{{\!{\rm d}^dy\,}}
\def\lV{{\overline V}}
\def\rd{{\rm d}}
\def\re{{\rm e}}

\begin{document}

\thesaurus{02(03.13.4;03.13.6;12.03.1;12.03.3)}

\titlerunning{On non--Gaussianity in the Cosmic Microwave Background}

\title{On non--Gaussianity in the Cosmic Microwave Background}

\authorrunning{D. Novikov, J. Schmalzing, and V. Mukhanov}

\author{
Dmitri Novikov\inst{1,2,3,}\thanks{novikov@theorie.physik.uni-muenchen.de},
Jens Schmalzing\inst{1,4,}\thanks{jens@theorie.physik.uni-muenchen.de}, and
Viatcheslav F.\ Mukhanov\inst{1,}\thanks{mukhanov@theorie.physik.uni-muenchen.de}
}

\makeatletter
\if@referee
\titlerunning{On non-Gaussianity in the CMB}
\authorrunning{Novikov, Schmalzing, Mukhanov}
\fi
\makeatother

\institute{
Ludwig--Maximilians--Universit\"at, 
Theresienstra{\ss}e 37, 
80333 M\"unchen, Germany.
\and
Astronomy Department,
Keble Road,
Oxford OX1 3RH,
United Kingdom.
\and
Astro-Space Center,
P.\ N.\ Lebedev Physical Institute,
84/32 Profsoyuznaya st.,
Moscow, 117810,
Russia.
\and
Teoretisk Astrofysik Center,
Juliane Maries Vej 30,
DK--2100 K{\o}benhavn {\O},
Denmark.
}

\mail{Jens Schmalzing}

\offprints{Jens Schmalzing}

\date{Version of September 11, 2000.  Accepted for publication in Astronomy and Astrophysics.}

\maketitle

\begin{abstract}
We consider a cosmological model with non--Gaussian initial
perturbations, which in principle could be generated in non--standard
inflationary scenarios with two or more scalar fields.  In particular
we focus our attention on the model proposed by
{\citet{linde1997}}, where perturbations are quadratic in a
Gaussian field.  These perturbations, if they exist, have to be
observable as a non-Gaussian distribution of the CMB signal on the
sky.  In order to efficiently pick up the non--Gaussian signal in CMB
maps of degree resolution, one can use Minkowski Functionals and peak
statistics.  Our paper contains the theoretical predictions of the
properties for Minkowski Functionals and distributions of peaks of the
CMB anisotropy in the model with "squared" Gaussian statistics.
Likelihood comparison of the four--year COBE DMR data to this
non--Gaussian model and the standard Gaussian model does not select
any of them as most likely.

We also suggest an efficient algorithm for fast simulation of CMB maps
on the whole sky.  Using a cylindrical partition of the sphere, we
rewrite the spherical harmonics analysis as a Fourier transform in
flat space, which makes the problem accessible to numerically
advantageous FFT methods.
\end{abstract}


\section{Introduction}
\label{sec:introduction}

Observations of the Cosmic Microwave Background provide fundamental
information about the primordial inhomogeneities in the Universe.  The
temperature anisotropies on scales above a few degrees preserve
information about primordial density fluctuations on scales larger
than the acoustic horizon at the moment of recombination.  Since the
first detection of the CMB anisotropy by the COBE satellite
{\citep{smoot1992,bennett1996}} several groups have
also reported observations on angular scales of about $1\degr$ which
roughly corresponds to the horizon size at recombination.  Obtaining
the spectrum of the primordial anisotropies on smaller scales with
high precision will allow us to verify several robust predictions of
inflation and permit to determine the values of important cosmological
parameters with high accuracy.  Future experiments
{\citep{bennett1995,bersanelli1996}} will yield all--sky maps
of the CMB with sufficiently high resolution and sensitivity.

Apart from estimates of the power spectrum with subsequent parameter
determination, the CMB data also provides very important constraints
on the nature of the perturbations that led to the formation of
large--scale structure {\citep{bond1998}}.  To extract this
information beyond the power spectrum we need to consider not
only the amplitudes, but also the phases of the temperature field.
One possible approach is under consideration in our paper.

Generally speaking, cosmological models based on the inflationary
paradigm predict adiabatic Gaussian fluctuations with a power spectrum
slightly different from the scale invariant one
({\citealt{mukhanov1981}}; see also
{\citealt{starobinsky1982,hawking1982,guth1985,bardeen1983}}).  The
Gaussianity of the density perturbations directly translates into
Gaussianity of the CMB temperature fluctuations on the sky.

However, along with standard inflationary models, at present there
still exist theories which are compatible with observations and
predict non--Gaussian primordial fluctuations.  Among them are, for
instance, non--standard inflationary scenarios with two or more scalar
fields, where one could expect significant deviations from Gaussianity
{\citep{linde1997,antoniadis1997,peebles1999:II,peebles1999:I}},
namely, the perturbations there are quadratic in a Gaussian field.
Another possibility to accumulate non--Gaussianity in the CMB signals
exists after recombination.  Even if the fluctuations were Gaussian at
the surface of the last scattering they may have acquired
non--Gaussian contributions due to subsequent weak gravitational
lensing {\citep{fukushige1995,bernardeau1997}} and due to various
foregrounds like dust emission, synchrotron radiation, or unresolved
point sources, to mention just a few {\citep{banday1996}}.
One should also take into account cosmic variance and additional
non--Gaussian instrumental noise in the observational data
{\citep{tegmark1997}}.

Therefore, establishing the Gaussian nature of the signal or detecting
some distinctive non--Gaussianity is crucial.  In fact, it would allow
us to reveal the nature of the primordial fluctuations.  For instance,
a confirmation that the fluctuations of the CMB are Gaussian would
leave practically no alternative to standard inflation, since it would
definitely rule out most of the models that predict non--Gaussian
fluctuations and are at present still compatible with observations.
In addition, the investigation of non--Gaussianity is also important
for a practical reason.  Most of the current techniques for estimating
the power spectrum from the observed signal with significant noise are
optimised for the Gaussian fields only
{\citep{feldman1994,knox1998,ferreira1998}}.

Many authors have searched for non--Gaussian signatures in CMB data
and in the large--scale structure using such diverse tests as peak
statistics {\citep{bardeen1986:bbks,bond1987,vittorio1987}}, the genus
curve {\citep{coles1988,smoot1994}}, higher--order correlations
{\citep{luo1993}}, peak correlations {\citep{kogut1996:tests}}, and
global Minkowski functionals
{\citep{gott1990,winitzki1997,schmalzing1998}}.

The first analysis of two--dimensional theoretical maps of the
temperature fluctuations that used the total area, length of the
boundary and genus for the excursion set was done by
{\citet{gott1990}} although without referring to Minkowski
functionals.  {\citet{schmalzing1998}} discuss the application of the
Minkowski functionals to all--sky maps taking into account the
curvature of the celestial sphere.  They apply these statistics to the
high--latitude portion of the four--year COBE DMR data and argued for
its advantage as a test of Gaussian signal.  They conclude that the
field is consistent with a Gaussian random field.
{\citet{colley1996}} measure the genus of the temperature fluctuations
in the COBE DMR 4-year sky maps and come to a similar conclusion.
{\citet{heavens1999}} compute the bispectrum of the four--year COBE
datasets and again fail to find evidence for non--Gaussian behaviour.

However, {\citet{ferreira1998}} study the distribution of an estimator
for the normalised bispectrum and conclude that Gaussianity is ruled
out at the confidence level at least of 99\%.  In a recent paper,
{\citet{novikov1999:minkowski}} suggest to use the partial Minkowski
functionals as quantitative descriptors of the geometrical properties
of CMB maps.  They apply their technique to the four--year DMR COBE
data corrected for the Galaxy contamination and also find significant
deviations from Gaussianity.  It was later shown {\citet{banday2000}}
that the observed non--Gaussianity can be explained by systematics.

While these findings may appear confusing at first sight
{\citep{bromley1999}}, they actually highlight the importance of
studying non--Gaussian fields in a quantitative way.  After all, while
the Gaussian random field is a well--defined notion, non--Gaussian
fields are still have to be substantiated.  For instance,
{\citet{kogut1996:tests}} introduce a non-Gaussian model where
perturbations in $\ell$--space are drawn from independent
$\chi_{\nu}^2$ distributions with $\nu$ degrees of freedom, while
{\citet{white1999}} puts forward the use of $\chi_{\nu}^2$
distributions in real rather than in $\ell$--space.

In our paper, we consider a cosmological model with non--Gaussian
initial conditions, with a statistics in real space following a
$\chi^2$ distribution with one degree of freedom.  This type of
perturbations results in very special non--Gaussianity of the CMB
anisotropy.  Therefore, it is well suited for the study of statistical
methods.  We calculate the Minkowski functionals and the distribution
functions of extrema of the CMB field smoothed with COBE resolution in
this model, and compare the outcome to the results for a Gaussian
field with the same spectrum.

This article is organised as follows.  In Section~\ref{sec:theory} we
briefly review the definition of the Minkowski functionals and the
distribution of extrema in two dimensions, and compare their
expectation values for a Gaussian field and a $\chi^2$ field with the
same smoothing scale and spectrum.  As an aside, we investigate the
behaviour of the one--point probability distribution under smoothing.
In Section~\ref{sec:data} we first outline an algorithm for doing Fast
Fourier Transforms on the sphere that was used for our CMB map
simulations.  Afterwards, we compare the Minkowski functionals and the
peak statistics of the COBE DMR data to both the Gaussian and the
non--Gaussian model.  Section~\ref{sec:outlook} summarises and
provides an outlook.

\section{Theory}
\label{sec:theory}

\subsection{Gaussian and $\chi^2$ random fields}
\label{sec:random}

Let us consider the temperature fluctuations $\frac{\Delta{T}}{T}$ of
the CMB temperature on the sky, parametrised with spherical
coordinates $\vartheta$ and $\varphi$.  Normalising with respect to
the dispersion $\sqrt{\sigma}$, where
$\sigma=\left\langle\left(\frac{\Delta{T}}{T}\right)^2\right\rangle$
is the variance of the temperature fluctuations, we obtain a random
field $u(\vartheta,\varphi)$ with zero mean and unit variance:
$\langle{u}\rangle=0$, $\langle{u^2}\rangle=1$.

In the following, we will consider two models for the random field
given by the normalised temperature fluctuations on the sky.

The standard is to model $u$ as a Gaussian random field.  The
properties of Gaussian random fields are very well known (see e.g.\
{\citealt{adler1981}}).

Apart from that, we will also use a $\chi^2$ field with one degree of
freedom, as suggested by the model of {\citet{linde1997}}.  In
order to retain zero mean and unit variance, we use the relation
\begin{equation} \psi=\frac{1-\phi^2}{\sqrt{2}} \label{eq:square}
\end{equation} to calculate a realisation $\psi$ of a $\chi^2$ field
from a given realisation $\phi$ of a Gaussian random field.

\subsection{Minkowski functionals and peak statistics}
\label{sec:minkowski}

\begin{figure}
\includegraphics[width=\linewidth]{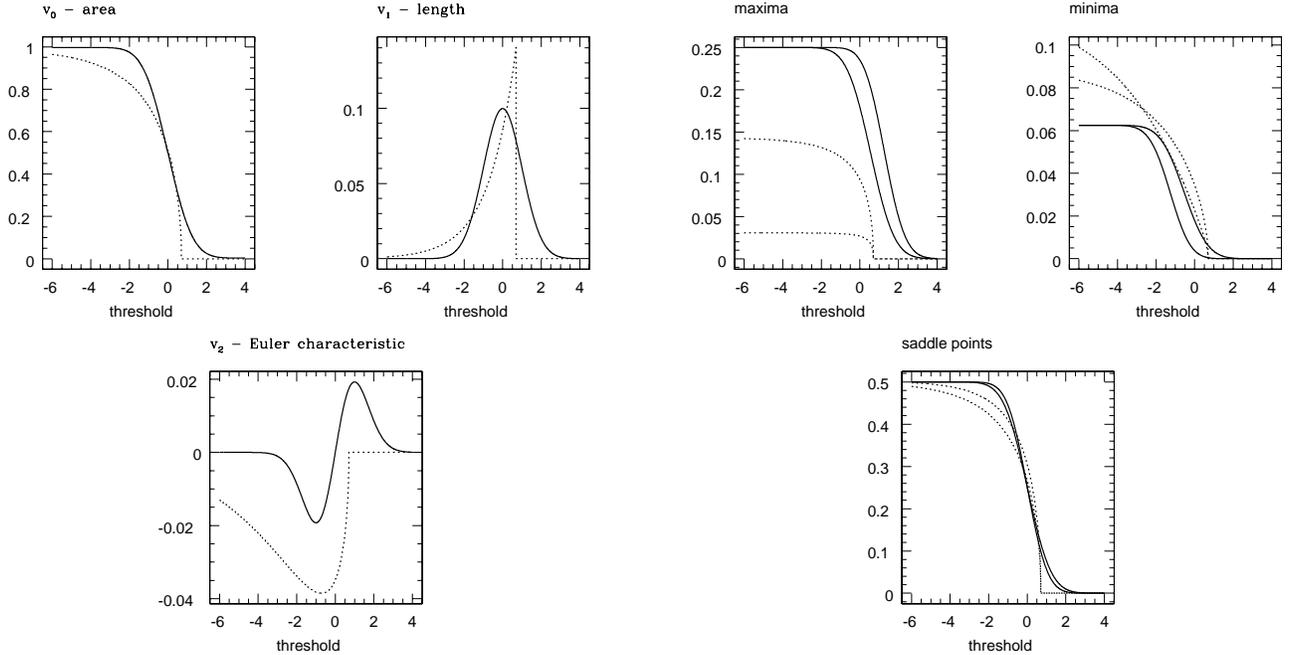}
\caption{
\label{fig:anal_minkowski}
The analytical expectation values of the Minkowski functionals for the
Gaussian field (solid) and the $\chi^2$ field (dashed).  }
\end{figure}

\begin{figure}
\includegraphics[width=\linewidth]{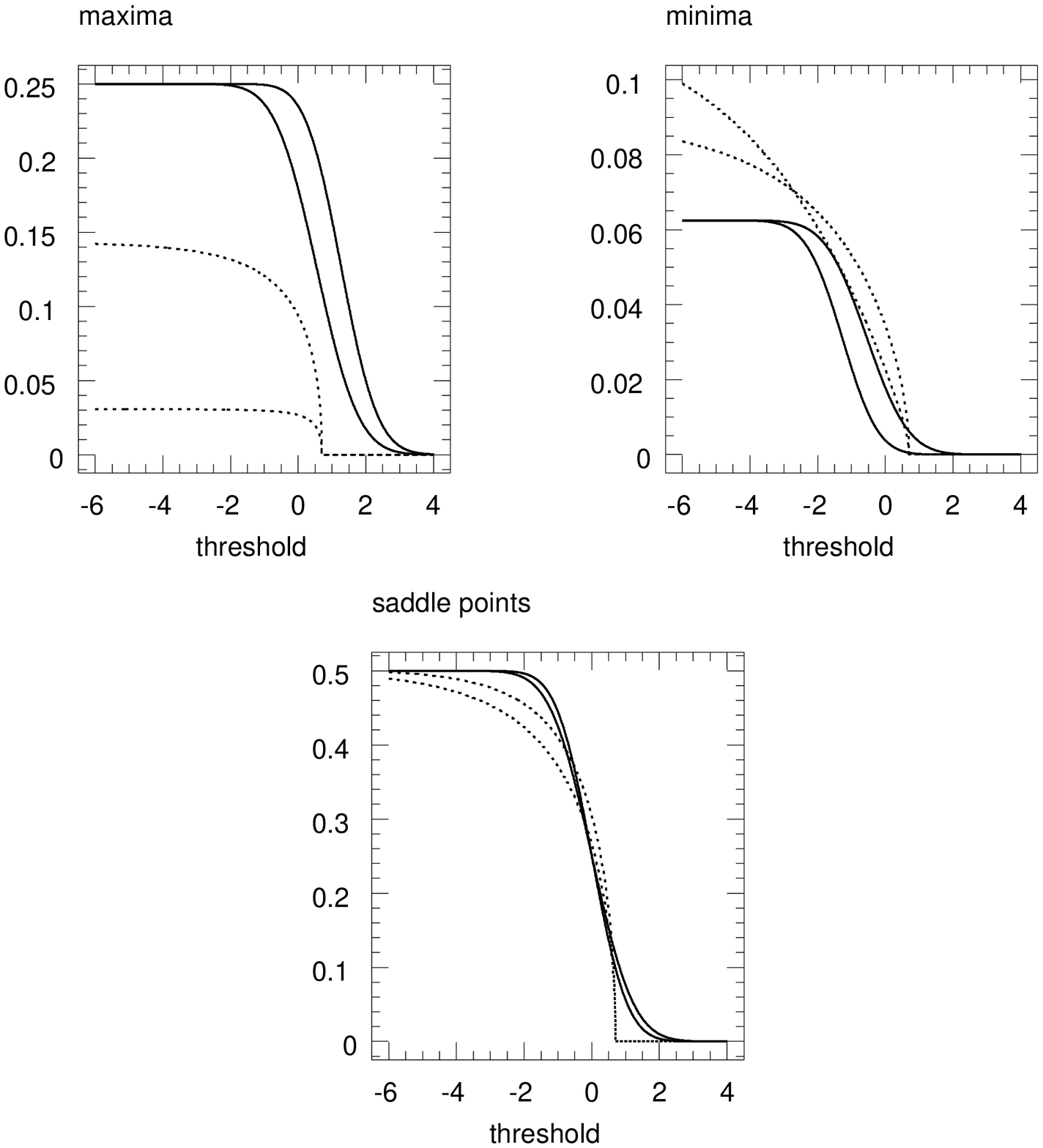}
\caption{
\label{fig:anal_extrema}
The analytical expectation values of the extrema distributions for the
Gaussian field (solid) and the $\chi^2$ field (dashed).  }
\end{figure}

Choosing a threshold $\nu$, we can divide the sphere into two parts:
Hot regions where the random field $u$ passes the threshold, and cold
regions where $u<\nu$.  The hot region is also called the excursion
set of the field $u$ over the threshold $\nu$.  Its properties can be
characterised by a large variety of geometrical quantities, including
the Minkowski functionals {\citep{minkowski1903}}.

From the mathematical point of view, Minkowski functionals have
several distinguishing properties.  {\citet{hadwiger1957}}
showed that all global morphological descriptors (satisfying
translational invariance and additivity) for patterns in
$d$-dimensional space are linear combinations of just $d+1$ Minkowski
functionals $v_\mu(\nu)$, with $\mu\in\{0,\ldots,d\}$.  Moreover, in
up to four dimensions all Minkowski functionals have simple and
intuitive geometrical meanings.  In two dimensions these are:

\begin{description}
\item[{\bf Area:}] $v_0(\nu)$ is the total area of all hot regions,
that is points where $u(\vartheta,\varphi)>\nu$.
\item[{\bf Boundary length:}] $v_1(\nu)$ is proportional to the total
length of the boundary between cold and hot regions.
\item[{\bf Euler characteristic:}] $v_2(\nu)$, being a purely
topological quantity, counts the number of isolated hot regions minus
the number of isolated cold regions.  This is not exactly true on the
sphere, where a generalised Gauss--Bonnet theorem holds
{\citep{allendoerfer1943}}, but nevertheless we will use this
relation in the following calculations.
\end{description}

Since Minkowski functionals are additive with respect to isolated
parts of the sky, they can be used for patchy or incomplete coverage.
Their calculation requires only $O(N)$ operations for calculating the
Minkowski functionals on a map of $N$ pixels, once the first and
second derivatives of the field at the pixel locations are known
{\citep{schmalzing1998}}.

Both for a Gaussian and a $\chi^2$ random field, the Minkowski
functionals in two dimensions are known analytically
{\citep{tomita1990,worsley1994,schmalzing1999:diss}}.  Since the
field is normalised to unit variance, the analytical values depend on
a single parameter $\tau=\langle{u_{,i}^2}\rangle$, given as the
variance of any of the field's first derivatives.  Note that $\tau$
has dimensions of inverse length squared, so one usually interprets
$r_{\text{corr}}=(2\tau)^{-1/2}$ as the so--called ``correlation
length'' of the random field $u$.

For the Gaussian random field, we
have\footnote{$\Phi(x)=\frac{2}{\sqrt{\pi}}\int_0^x\dt\re^{-t^2}$ is
the error function.}
\begin{equation}
\begin{split}
v_0(\nu)
&=
\frac{1}{2}-\frac{1}{2}\Phi\left(\frac{\nu}{\sqrt{2}}\right),
\\
v_1(\nu)
&=
\frac{\sqrt{\tau}}{8} \exp\left(-\frac{\nu^2}{2}\right),
\\
v_2(\nu)
&=
\frac{\tau}{\sqrt{8\pi^3}} \nu\exp\left(-\frac{\nu^2}{2}\right).
\end{split}
\end{equation}

For the $\chi^2$ random field with one degree of freedom we
have\footnote{The results for an arbitrary number of degrees of
freedom can be found in {\citet{schmalzing1999:diss}}.}
\begin{equation}
\begin{split}
v_0(\nu)
&=
\Phi\left(\sqrt{\frac{1}{2}-\frac{\nu}{\sqrt{2}}}\right),
\\
v_1(\nu)
&=
\frac{\sqrt{\tau}}{4\sqrt{2}}
\exp\left(-\frac{1}{2}+\frac{\nu}{\sqrt{2}}\right),
\\
v_2(\nu)
&=
\frac{\tau}{4\pi\sqrt{\pi}} \sqrt{\frac{1}{2}-\frac{\nu}{\sqrt{2}}}
\exp\left(-\frac{1}{2}+\frac{\nu}{\sqrt{2}}\right).
\end{split}
\end{equation}

Figure~\ref{fig:anal_minkowski} shows a comparison of the expectation
values of the Minkowski functionals of the Gaussian and $\chi^2$
random fields for the same parameter $\tau$.

By the Morse theorem {\citep{morse1969}}, the Euler
characteristic $\chi(\nu)$ of the excursion set above a certain
threshold $\nu$ is related to the number of extrema above this
threshold by
\begin{equation}
\chi(\nu)=N_{\text{max}}(\nu)+N_{\text{min}}(\nu)-N_{\text{sad}}(\nu).
\label{eq:morse}
\end{equation}
Here $N_{\text{max}}(\nu)$, $N_{\text{min}}(\nu)$ and
$N_{\text{sad}}(\nu)$ denote the number of maxima, minima and saddle
points, respectively, where the value of the field itself lies above
the threshold $\nu$.  Along with the Minkowski functionals, we use
these three distribution functions of all possible kinds of extrema --
maxima, minima and saddle points -- as another measure of
non--Gaussianity.  We always normalise these numbers by the total
number
$N_{\text{ext}}=N_{\text{max}}(-\infty)+N_{\text{min}}(-\infty)+N_{\text{sad}}(-\infty)$
of extrema in the field, giving number densities
$n_{\text{max}}(\nu)$, $n_{\text{min}}(\nu)$ and
$n_{\text{sad}}(\nu)$, respectively.

The peak statistics for the Gaussian random field have been thoroughly
investigated {\citep{bond1987,bardeen1986:bbks}}, but unfortunately
cannot be written in a simple form.  The same holds for the $\chi^2$
field, whose extrema distribution functions are in principle easily
obtained from the Gaussian case via Equation~(\ref{eq:square}).
Furthermore, the shapes of the Minkowski functional curves are
universal, that is they do not depend on anything but the statistical
nature (Gaussian or $\chi^2$) of the underlying random field, and the
analytical expectation values only contain constants of
proportionality, parametrised by the correlation radius
$r_{\text{corr}}$.  This is not the case for the distribution
functions of extrema, where a second parameter $\gamma$ also affects
the shape of the curves.  We demonstrate this in
Figure~\ref{fig:anal_extrema}, by showing the expectation values for
two different values of $\gamma$.

Since the Euler characteristic is proportional to the third Minkowski
functional $v_2$, the distribution functions of extrema and the
Minkowski functionals are not independent.  In the following
Section~\ref{sec:data}, we will use the first two Minkowski
functionals $v_0$ and $v_1$, and the three distribution functions of
extrema to test our models against the four--year COBE DMR data.
However, let us first take a look at the effects of smoothing on
non--Gaussian fields.

\subsection{Non--Gaussianity and Gaussian smoothing}
\label{sec:smoothing}

\begin{figure*}
\includegraphics[width=.33\linewidth]{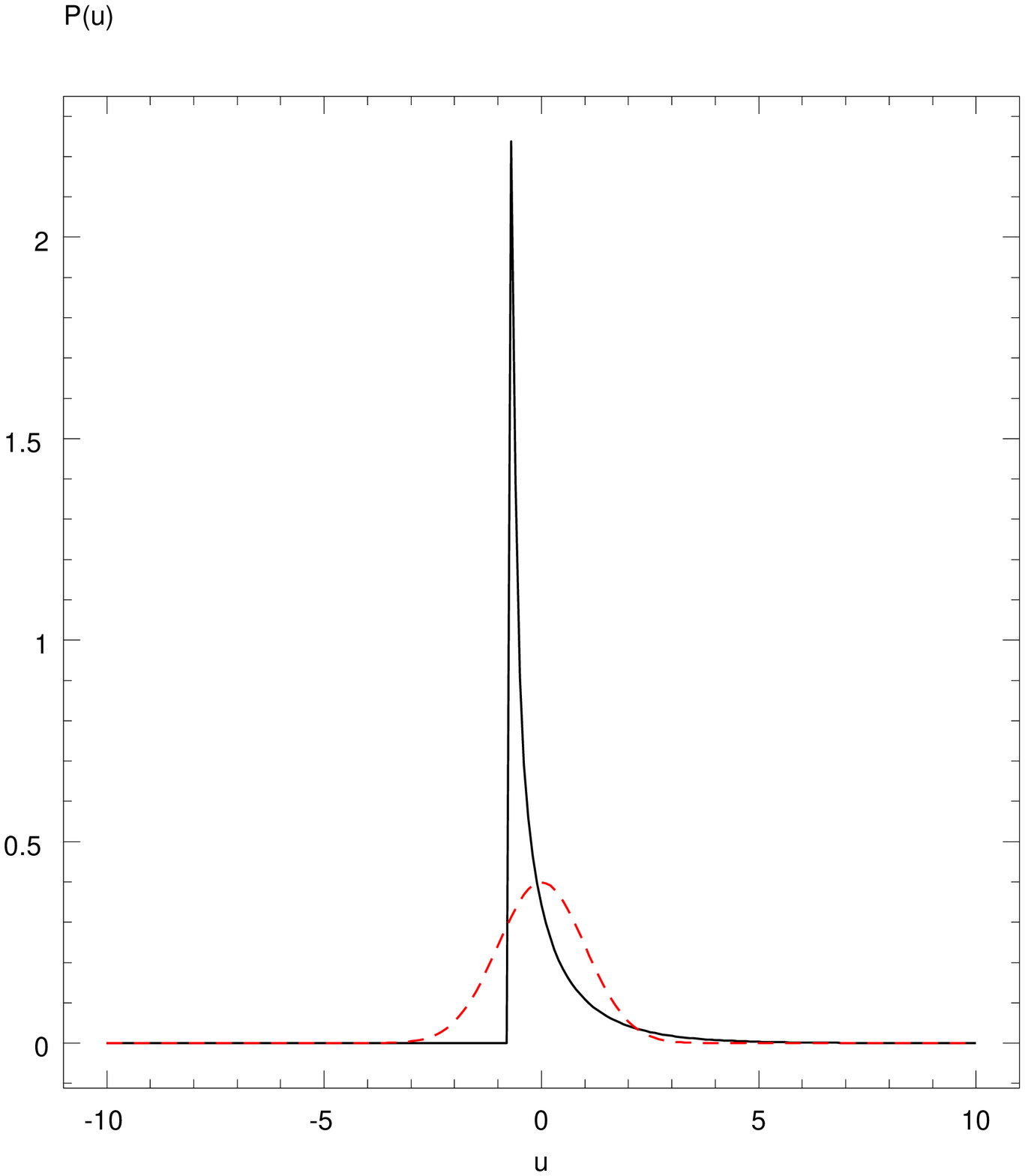}\hfill
\includegraphics[width=.33\linewidth]{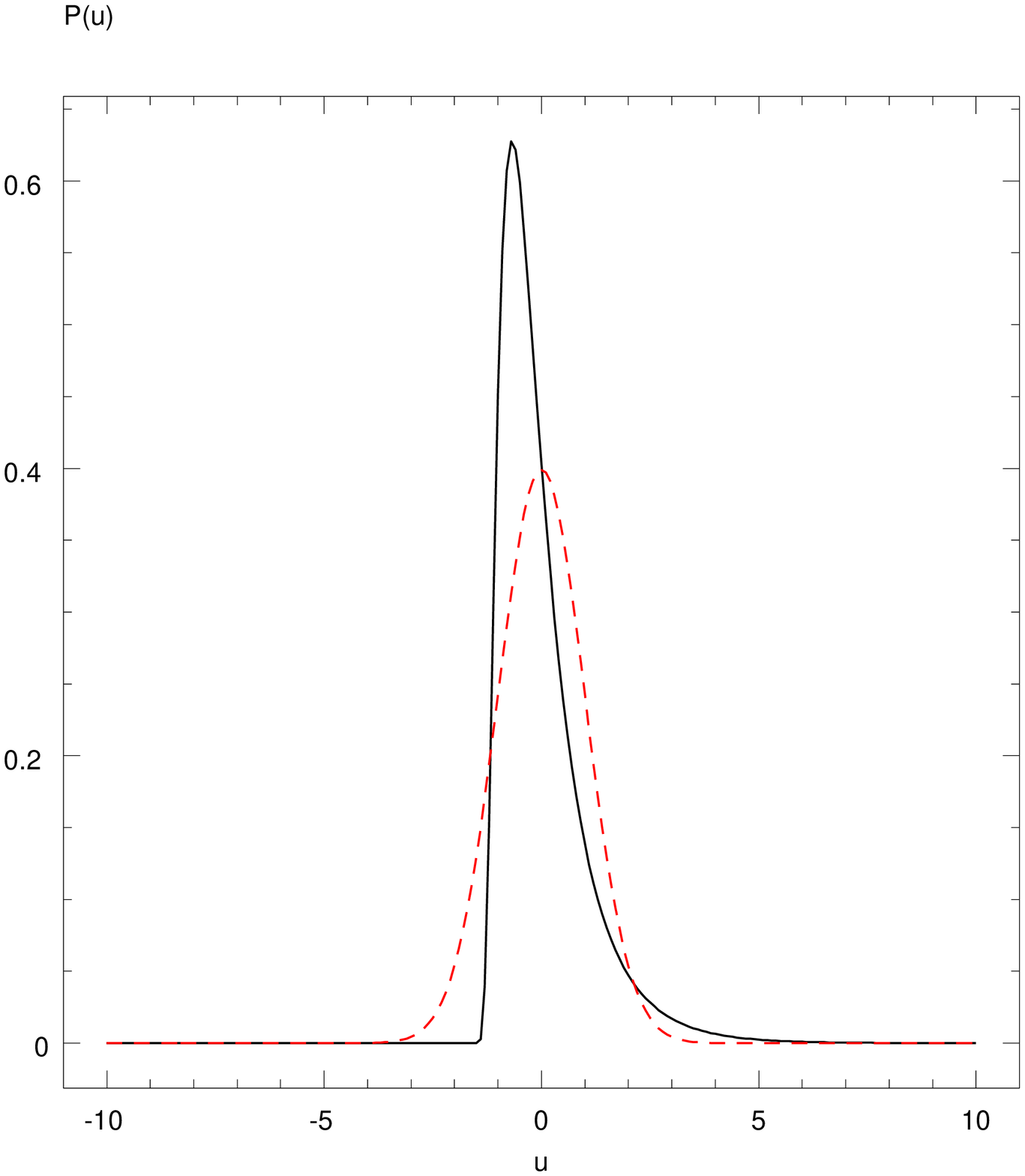}\hfill
\includegraphics[width=.33\linewidth]{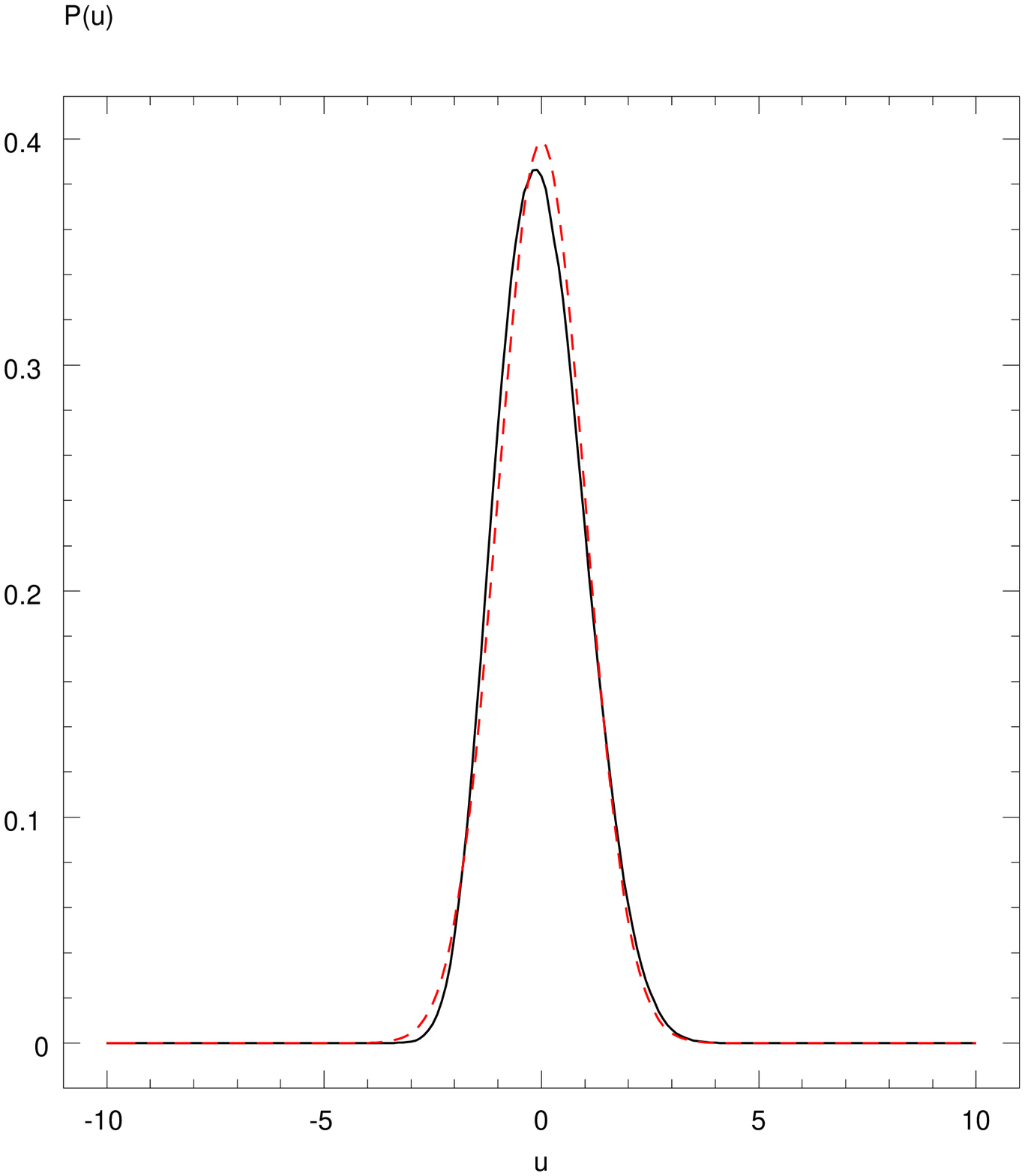}
\caption{
\label{fig:smoothing}
The PDFs of a smoothed $\chi^2$ random field (solid lines).  From left
to right, the smoothing scale increases, ranging from no smoothing
(left panel) over little smoothing (middle panel) to considerable
smoothing (right panel).  The transition of the curve's shape from
clearly non--Gaussian to almost Gaussian is obvious.  For comparison,
the Gaussian bell curve is plotted in all three examples (dashed
lines).  }
\end{figure*}

As an aside, we note that CMB temperature maps usually suffer from
considerable smearing through the finite instrument beam size and are
usually smoothed in order to reduce the noise level.  Therefore, a
non--Gaussian CMB signal may appear in the map with a much distorted
distribution.  Here, we attempt to quantify this effect.  We focus on
the one--point probability distribution (PDF) of the field, because it
is easy to handle and also closely related to the zeroth Minkowski
functional, being nothing but its nagative first derivative.  It is of
course conceivable that a non--Gaussian random field still has a
Gaussian PDF.  Other statistics can in principle be treated in a
similar fashion.

We start from the cosmological CMB signal and consider it as a random
field $u(\bx)$ on some $d$--dimensional space, where in our case
$d=2$.  Introducing a smoothing filter $g(\bx,t)$, where $t$ is the
smoothing scale and $g(\bx,0)=\delta(\bx)$ -- since the unsmoothed
field should be equal to the original field -- we obtain the smoothed
field $u(\bx,t)$ by convolution:
\begin{equation}
u(\bx,t)=\cN(t)\int\ddy g(\bx-\by,t)u(\by),
\label{eq:smooth}
\end{equation}
where the constant $\cN(t)$ is chosen such that the smoothed field
remains normalised to unit variance.  We may think of $u(\bx,t)$ as a
field in $d+1$--dimensional ``scale space'' {\citep{koenderink1991}}.

If the filter $g$ is Gaussian, it obeys the diffusion equation
\begin{equation}
\frac{\partial{g(\bx,t)}}{\partial{t}}=t{\Delta}g(\bx,t),
\label{eq:diffusion}
\end{equation}
where the Laplacian $\Delta$ is of course taken with respect to the
spatial coordinates $\bx$ only.

Combining Equations~(\ref{eq:smooth}) and (\ref{eq:diffusion}), we can
write down an ``evolution equation'' for the field $u$ as the scale
$t$ changes:
\begin{equation}
\frac{\partial{u(\bx,t)}}{\partial{t}}
=
t\left(\Delta+r_{\text{corr}}^{-2}\right)u(\bx,t).
\label{eq:field}
\end{equation}
The second expression $r_{\text{corr}}^{-2}$ in the operator enters
through the scale--dependence of the normalisation factor $\cN(t)$.
Details of the calculation can be found in
Appendix~\ref{sec:equation-field}.

This equation enables us to study the one--point probability
distribution $P(u,t)$ of the smoothed field $u(\cdot,t)$ as the
smoothing scale changes.  Writing this probability density as
\begin{equation}
P(u,t)=\left\langle\delta(u(\bx,t)-u)\right\rangle
\label{eq:definition}
\end{equation}
and applying the partial derivative with respect to $t$, we
immediately obtain
\begin{equation}
\frac{\partial{P(u,t)}}{\partial{t}}
=
-t\frac{\partial}{\partial{u}}\left[
\left(\langle\Delta{u}\rangle_u+\frac{u}{r_{\text{corr}}^2}\right)
P(u,t)\right].
\label{eq:probability}
\end{equation}
We refer the reader to Appendix~\ref{sec:equation-probability} for the
technical details of this calculation.  The quantity
$\langle\Delta{u}\rangle_u$ denotes the average of the Laplacian of
the field $u(\bx,t)$ under the condition that its value $u$ is fixed.
It is interesting to note that this equation is written in
conservative form, that is its integral over $\du$ vanishes.

For a variety of random fields, the conditional average of the
Laplacian can be calculated analytically.  Most notably, for both the
Gaussian and the $\chi^2$ field
$\langle\Delta{u}\rangle_u=-\frac{u}{r_{\text{corr}}^2}$.  In these
cases, the right--hand side of Equation~(\ref{eq:probability}) simply
becomes equal to zero.

As far as the Gaussian random field is concerned,
$P(u)\propto\exp(-u^2/2)$ is a stationary solution of
Equation~(\ref{eq:probability}).  This reflects the well--known fact
that a Gaussian random field stays Gaussian under smoothing, and in
fact under any linear filtering.

In the case of the $\chi^2$ field, however, the probability
distribution is non--zero only for $u<1/\sqrt{2}$ by
Equation~(\ref{eq:square}), and is not differentiable at the upper
bound of the field.  Therefore, under smoothing the field evolves away
from the $\chi^2$ distribution, and finally ends up with a
distribution close to the stationary Gaussian solution at large
smoothing lengths.  Figure~\ref{fig:smoothing} illustrates the effect
using a $\chi^2$ random field with various degrees of smoothing.

This shows that although our two models are fairly different at first
glance, the noise, beam size and other effects of actual observations
tend to make the fields appear more similar.  We shall look upon the
practical consequences of this now, using the four--year COBE DMR data
as an example.

\section{Data analysis}
\label{sec:data}

\begin{figure}
\includegraphics[width=\linewidth]{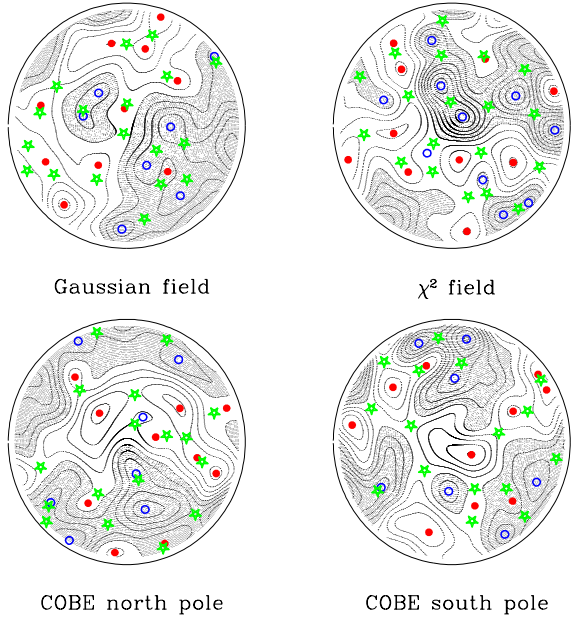}
\caption{
\label{fig:circles}
Four maps showing realisations of the Gaussian and $\chi^2$ random
fields and the northern and southern parts of the COBE DMR map,
respectively.  All maps extend from the pole to $30\degr$ latitude.
The lines show isocontours of the field at half integer multiples of
the standard deviation.  Shaded areas indicate regions where the field
value drops below the mean.  Open circles show the positions of
minima, full circles correspond to maxima, and saddle points are
indicated by drude's feet.  }
\end{figure}

We illustrate our methods on the COBE DMR four--year all--sky maps
{\citep{bennett1996}}.  We work with the map constructed from all
three bands and both channels of the DMR instrument by the so--called
subtraction method.  Although this method is supposed to remove most
of the galactic contamination, we avoid the galactic plane completely
by applying a cut to $45\degr$ latitude.

For the statistical tests presented in the following, we use a large
number of realizations of the Gaussian and $\chi_1^2$ random fields.
In both cases, we calculate the fields using a scale--free spectrum
with a correlation radius of $1\degr$.  To mimic the DMR beam, we
smooth the fields with a Gaussian filter of $7\degr$ FWHM.  Examples
of realisations of the Gaussian and the $\chi^2$ random fields,
together with the COBE DMR maps, are shown in
Figure~\ref{fig:circles}.

\subsection{Fast Fourier Transform on the sphere}
\label{sec:fft}

\begin{figure}
\includegraphics[width=\linewidth]{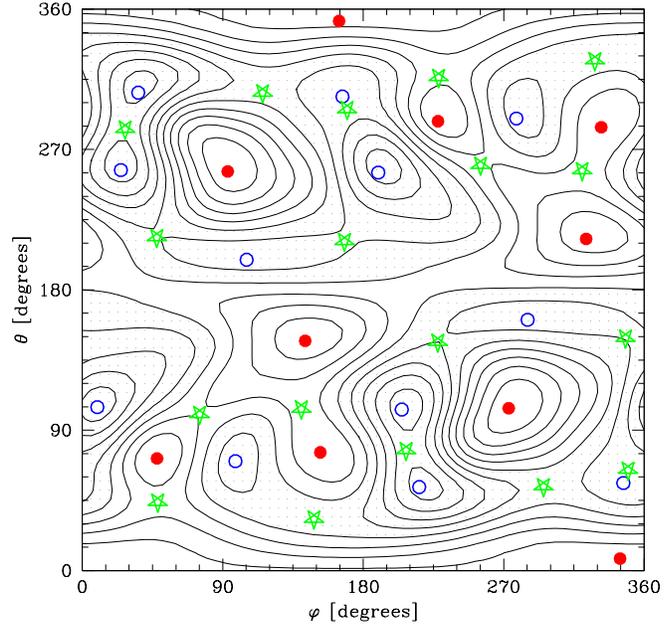}
\caption{
\label{fig:squaremap}
A COBE all--sky map, projected onto the plane in cylindrical
coordinates and mirrored at the North pole to give a square suitable
for the FFT method described in the text.  The meaning of the various
line and symbol styles is explained in the caption of
Figure~\protect~\ref{fig:circles}.  }
\end{figure}

Statistical tests on CMB maps usually involve simulating large numbers
of model maps from given power spectra.  The na{\"\i}ve approach to
this computational task is extremely time--consuming, requiring
$\cO(N^2M^2)$ operations, where $N$ and $M$ are the number of pixels
on the equator and the number of multipoles, respectively.  Here we
use a Fast Fourier Transform (FFT) in flat space over spherical
coordinates outlined by {\citet{risbo1996}}.  This means that the FFT
is used in both the polar and the azimuthal direction, while other
methods usually perform the FFT over the polar angle alone.  Let us
first outline the mathematical and computational ideas.

Given a realisation of a random field in terms of its spherical
harmonics coefficients $a_{{\ell}m}$, we wish to calculate its values
$u(\vartheta,\varphi)$ in real space parametrised with spherical
coordinates $(\vartheta,\varphi)$.  This can be done through the
spherical harmonics expansion
\begin{equation}
u(\vartheta,\varphi)
=
\sum_{\ell=0}^{M-1}\sum_{m=-\ell}^{+\ell}
a_{{\ell}m}Y_{{\ell}m}(\vartheta,\varphi),
\label{eq:harmonicseries}
\end{equation}
where the functions $Y_{{\ell}m}$ are the spherical harmonics.  They are
related to the associated Legendre functions $P_{{\ell}m}$ by
\begin{equation}
Y_{{\ell}m}(\vartheta,\varphi)
=
h_{{\ell}m}P_{{\ell}m}(\cos\vartheta)\re^{im\varphi},
\end{equation}
where $m$ and $\ell$ are called the order and the degree of the
function, respectively, and
\begin{equation}
h_{{\ell}m}=\sqrt{\frac{2\ell+1}{4\pi}\frac{(\ell-m)!}{(\ell+m)!}}.
\end{equation}

All spherical harmonics $Y_{{\ell}m}$ enjoy the symmetry property
\begin{equation}
Y_{{\ell}m}(\vartheta,\varphi)=Y_{{\ell}m}(2\pi-\vartheta,\pi+\varphi).
\end{equation}
This can be used to extend the field $u(\vartheta,\varphi)$, which is
normally defined for $(\vartheta,\varphi)\in[0,\pi]\times[0,2\pi]$, to
the whole square $[0,2\pi]^2$, simply by mirroring the field at the
point $(\pi,\pi)$.  The resulting square map is illustrated by an
example in Figure~\ref{fig:squaremap}.

The extended field in real space, which now consists of two identical
copies of the actual map, is periodic both in $\vartheta$ and in
$\varphi$, and can therefore be expanded into a real--space Fourier
series:
\begin{equation}
u(\vartheta,\varphi)
=
\sum_{j=0}^{M-1}\sum_{k=0}^{M-1}b_{jk}\re^{ij\vartheta+ik\varphi},
\label{eq:fourierseries}
\end{equation}

Inverting Equation~(\ref{eq:fourierseries}) and inserting
Equation~(\ref{eq:harmonicseries}) allows us to establish a relation
between the Fourier coefficients and the spherical harmonics
coefficients $a_{{\ell}m}$:
\begin{multline}
b_{jk}
=
\sum_{\ell=k}^{M-1}\left.a_{{\ell}m}h_{{\ell}m}p_{{\ell}mj}\right|_{m=k}
\\
+
\sum_{\ell=M-k}^{M-1}\left.a_{{\ell}m}h_{{\ell}m}p_{{\ell}mj}\right|_{m=k-M}
,
\label{eq:fouriercoefficients}
\end{multline}
where the $p_{{\ell}mj}$ are the coefficients of the Fourier series
of the associated Legendre functions:
\begin{equation}
P_{{\ell}m}(\cos\vartheta)
=
\sum_{j=0}^{M-1}\re^{ij\vartheta}p_{{\ell}mj}.
\end{equation}

Using the well--known {\citep{press1987}} recurrence relation for
the associated Legendre functions, it is easy to derive recurrence
relations for numerical evaluation of the $p_{{\ell}mj}$.

Let us now summarise the steps for efficiently calculating a CMB map
in real space from its spherical harmonics coefficients.
\begin{enumerate}
\item
Obtain a realisation of the random field in spherical harmonics
$a_{{\ell}m}$.
\item
Calculate the intermediate Fourier coefficients $b_{jk}$ using
Equation~(\ref{eq:fouriercoefficients}).  This step requires
$\cO(M^3)$ operations.
\item
Perform the two--dimensional Fast Fourier Transform from
Equation~(\ref{eq:fourierseries}) in order to obtain the real--space
realisation $u(\vartheta,\varphi)$.  This step requires
$\cO(N^2\log{M})$ operations.
\end{enumerate}

Obviously, our approach produces a cylindrical pixelisation in real
space, resulting in non--uniform pixels.  This makes it unsuitable for
applications that involve quadratures, where uniform pixelisations
(e.g.\ {\citealt{gorski1999}}) are desirable.  However, the method
is well suited for the calculations presented below, because it allows
large oversampling of the map in real space, that is $N\ll{M}$, with
little computational effort.  This is desirable both for accurate
calculations of Minkowski functionals of a map by tracing the contours
with high accuracy, and for determining the exact positions of
extrema.

\subsection{Minkowski functionals}
\label{sec:data_minkowski}

\begin{table}
\begin{center}
\begin{tabular}{llccccc}
&& $P_{\text{area}}$ & $P_{\text{length}}$
 & $P_{\text{max}}$ & $P_{\text{min}}$ & $P_{\text{sad}}$ \\
South & Gaussian   & 94\% & 12\% & 38\% & 52\% & 53\% \\
      & $\chi_1^2$ & 87\% & 61\% & 44\% & 56\% & 57\% \\
North & Gaussian   & 33\% & 44\% & 89\% & 87\% & 95\% \\
      & $\chi_1^2$ & 31\% & 28\% & 76\% & 81\% & 88\%
\end{tabular}
\end{center}
\caption{
\label{tab:probabilities}
Summary of the probabilities of obtaining the COBE DMR data as a
realisation of the models -- a Gaussian random field and a $\chi^2$
random field -- considered here.  The analysis is done separately for
the Northern and Southern polar caps.  Since we use five related
statistics to obtain these probabilities, five values are shown for
each case.  }
\end{table}

\begin{figure}
\includegraphics[width=\linewidth]{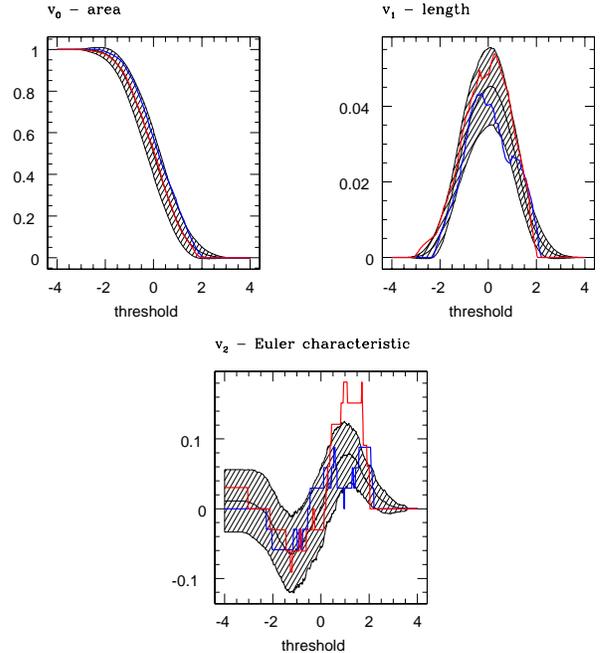}
\caption{
\label{fig:cobe_minkowski}
The Minkowski functionals of the COBE DMR data in comparison with the
average values for a Gaussian random field.  The areas indicate the
standard deviation.  }
\end{figure}

The Minkowski functionals of the COBE DMR maps are displayed in
Figure~\ref{fig:cobe_minkowski}.  Since the Northern and Southern
polar caps were analysed separately, we show two curves in each panel.
The shaded areas indicate the average and standard deviation of 1000
realisations of a Gaussian random field with the same two--point
characteristics as the COBE data.

In order to assess the probability that the COBE DMR maps are indeed a
realisation of our toy models, we use a non--parametric test
{\citep{novikov2000}}.

We computed $N=1000$ realisation of each random field.  The $n$th
realisation yields Minkowski functionals $V_\mu^{(n)}(\nu)$, with the
threshold $\nu$ ranging from $-5$ to $5$.  We calculate the average
Minkowski functionals
\begin{equation}
\lV_\mu(\nu)=\frac{1}{N}\sum_{n=1}^NV_\mu^{(n)}(\nu)
\end{equation}
for the random field, and determine the $L_1$--distance of each
realisation from this average:
\begin{equation}
\Delta_\mu^{(n)}=\int\rd\nu\left|V_\mu^{(n)}(\nu)-\lV_\mu(\nu)\right|.
\end{equation}
We then obtain the same quantity $\Delta_\mu^{\text{COBE}}$ from the
data, and test the hypothesis that this $\Delta_\mu^{\text{COBE}}$ is
distributed according to the distribution function of the
$\Delta_\mu^{(n)}$ of the model we are currently investigating.

The resulting probabilities are displayed in
Table~\ref{tab:probabilities}.  Note that we only show the values for
the first two functionals, the area $V_0$ and the length $V_1$, since
by Equation~(\ref{eq:morse}) the last Minkowski functional is a linear
combination of the extrema statistics analysed in
Section~\ref{sec:data_extrema}.

\subsection{Distribution of extrema}
\label{sec:data_extrema}

\begin{figure}
\includegraphics[width=\linewidth]{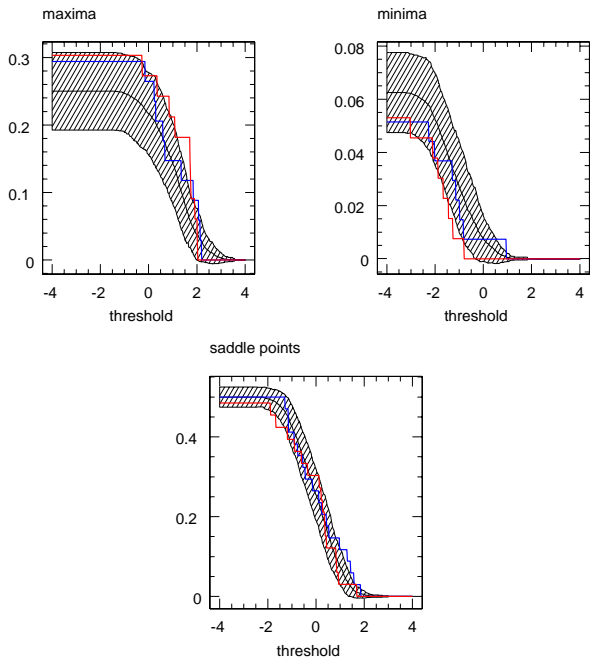}
\caption{
\label{fig:cobe_extrema}
The extrema statistics of the COBE DMR data in comparison with the
average values for a Gaussian random field.  The areas indicate the
standard deviation.  }
\end{figure}

For the COBE DMR data, the distribution functions of all three kinds
of extrema are shown in Figure~\ref{fig:cobe_extrema}.  For
comparison, we also show the mean and standard deviations over 1000
realisations of a Gaussian random field with the same two--point
characteristics.

In order to quantify the difference of the COBE data and the models,
we use a standard Kolmogorov--Smirnov test {\citep{press1987}}.
The resulting probabilities are shown in
Table~\ref{tab:probabilities}.

\section{Summary and Outlook}
\label{sec:outlook}

We have investigated a method for probing the non--Gaussianity of CMB
maps, based on Minkowski functionals and on distribution functions of
extrema.  In order to make the deviations from Gaussianity
quantitative, we used a model proposed by
{\citet{linde1997}}, producing to a $\chi^2$ distribution with
one degree of freedom for the primordial fluctuations.  Analytical
calculations for our statistics showed that the non--Gaussian and
Gaussian models produce markedly different results.

The situation may change, however, when smoothing is taken into
account.  We derived an equation governing the change of the
one--point probability distribution with the smoothing scale and found
that, while the Gaussian field remains Gaussian under smoothing, the
$\chi^2$ field evolves towards a more symmetric distribution with
closer resemblance to the Gaussian bell.  The outcome of this effect
became clear in an application of the method to the COBE DMR data --
we saw a slight preference of the Gaussian model, but this was by no
means significant.  Since all realisations of the maps were highly
oversampled in real space, the failure to discriminate between the two
models is definitely due to the rather large smoothing scale.
Therefore, an application of the method to the recently completed
\textsc{Boomerang} {\citep{bernardis2000}} and MAXIMA
{\citep{hanany2000}} experiments appears highly promising, as soon
as the data become available.

\section*{Acknowledgements}

We thank Benjamin Wandelt and Saleem Zaroubi for valuable comments.
JS wishes to thank Johannes Vahlensieck for interesting discussions.
DN acknowledges fellowships from the Alexander von Humboldt--Stiftung
and the Board of the Glasstone Benefaction.  This investigation was
supported by INTAS under grant number 97--1192 and by Danmarks
Grundforskningsfond through its support for TAC.  The COBE datasets
were developed by the NASA Goddard Space Flight Center under the
guidance of the COBE Science Working Group and were provided by the
NSSDC.


\appendix

\section{Derivation of Equation~(\ref{eq:field})}
\label{sec:equation-field}

We start from the smoothing prescription in
Equation~(\ref{eq:smooth}):
\begin{equation}
u(\bx,t)=\cN(t)\int\ddy g(\bx-\by,t)u(\by),
\end{equation}
and apply the partial derivative with respect to the smoothing scale
$t$.  Straightforward differentiation yields
\begin{multline}
\frac{\partial{u(\bx,t)}}{\partial{t}}
= \\
\frac{\partial{\cN(t)}}{\partial{t}}\int\ddy g(\bx-\by,t)u(\by)
+ \\
\cN(t)\int\ddy\frac{\partial{g(\bx-\by,t)}}{\partial{t}}u(\by).
\end{multline}
We use the diffusion equation~(\ref{eq:diffusion}) on the second
addend, and eliminate the integrals by reinserting the smoothing
prescription~(\ref{eq:smooth}).  This leads to:
\begin{equation}
\frac{\partial{u(\bx,t)}}{\partial{t}}
=
\frac{\frac{\partial{\cN(t)}}{\partial{t}}}{\cN(t)} u(\bx,t) +
t{\Delta} u(\bx,t).
\label{eq:partial}
\end{equation}
In order to evaluate the partial derivative of the normalisation
factor $\cN(t)$, we use the normalisation condition
\begin{equation}
1=\int\ddx u(\bx,t)^2
\label{eq:normal}
\end{equation}
and again apply the partial derivative with respect to $t$.  Together
with Equation~(\ref{eq:partial}), this becomes
\begin{equation}
\begin{split}
0
&=
\int\ddx u(\bx,t) \frac{\partial{u(\bx,t)}}{\partial{t}}
\\&=
\frac{\frac{\partial{\cN(t)}}{\partial{t}}}{\cN(t)}\int\ddx u(\bx,t)^2 +
t\int\ddx u(\bx,t){\Delta}u(\bx,t)
\\&=
\frac{\frac{\partial{\cN(t)}}{\partial{t}}}{\cN(t)}
-t\times r_{\text{corr}}^{-2}.
\end{split}
\end{equation}
The first integral is equal to one because of the normalisation
condition~(\ref{eq:normal}), while the second integral can be
rewritten as the variance of the gradient of $u(\bx,t)$ using partial
integration, which is equal to the inverse square of the correlation
radius $r_{\text{corr}}$ by definition.

Now we can eliminate the explicit reference to the normalisation
factor $\cN(t)$ from Equation~(\ref{eq:partial}) and obtain our final
result, Equation~(\ref{eq:field}):
\begin{equation}
\frac{\partial{u(\bx,t)}}{\partial{t}}
=
t\left(\Delta+r_{\text{corr}}^{-2}\right)u(\bx,t).
\end{equation}

\section{Derivation of Equation~(\ref{eq:probability})}
\label{sec:equation-probability}

This time, we investigate the properties of the one--point probability
distribution function $P(u)$ under smoothing.  We start from the
definition~(\ref{eq:definition}):
\begin{equation}
P(u,t)=\left\langle\delta(u(\bx,t)-u)\right\rangle.
\end{equation}

It is a common bad habit among physicists to use the same notation for
utterly different things.  We are no expection to this, employing the
same letter both for the smoothed random field and for the argument of
the probability distribution.  However, we always add the scale space
coordinate $(\bx,t)$ to the random field $u(\bx,t)$ , while the value
$u$ has no such argument.

At first, the partial derivative of this equation yields a functional
derivative of the $\delta$--function with respect to the field
$u(\bx,t)$.  Fortunately, it can be rewritten as an ordinary partial
derivative with respect to the value $u$.  Furthermore, the value $u$
(not the field $u(\bx,t)$!) only enters through the
$\delta$--function.  We can then insert Equation~(\ref{eq:field}) for
the evolution of the field under smoothing.  All in all, the
calculation proceeds as follows:
\begin{equation}
\begin{split}
\frac{\partial{P(u,t)}}{\partial{t}}
&=
\left\langle
\frac{\partial{\delta(u(\bx,t)-u)}}{\partial{u(\bx,t)}}
\times
\frac{\partial{u(\bx,t)}}{\partial{t}}
\right\rangle
\\&=
-\frac{\partial}{\partial{u}}
\left\langle
\delta(u(\bx,t)-u)\frac{\partial{u(\bx,t)}}{\partial{t}}
\right\rangle
\\&=
-t\frac{\partial}{\partial{u}}
\left\langle
\delta(u(\bx,t)-u)\left(\Delta+r_{\text{corr}}^{-2}\right)u(\bx,t)
\right\rangle
\\&=
-t\frac{\partial}{\partial{u}}
\left\langle
\delta(u(\bx,t)-u)\left({\Delta}u(\bx,t)+\frac{u}{r_{\text{corr}}^2}\right)
\right\rangle
\\&=
-t\frac{\partial}{\partial{u}}
\left[
\left(\langle\Delta{u}\rangle_u+\frac{u}{r_{\text{corr}}^2}\right)
P(u,t)\right],
\end{split}
\end{equation}
so we end up with Equation~(\ref{eq:probability}).

\end{document}